\documentclass[a4paper,1pt]{article}

\usepackage{fancyhdr}
\usepackage{multicol}
\usepackage{etoolbox}
\usepackage{cancel}
\usepackage[table]{xcolor}
 
\usepackage[font=small,labelfont=bf]{caption}
\DeclareCaptionFont{scriptsize}{\scriptsize}
\DeclareCaptionFont{tiny}{\tiny}
\usepackage{float}
\usepackage{amsmath, amssymb, setspace,cite,verbatim}
\usepackage{graphicx}
\usepackage{hyperref}
\usepackage{color}
\usepackage{wrapfig}

\usepackage{multicol}
\usepackage{etoolbox}
\usepackage{relsize}
\usepackage{multirow}

\usepackage{latexsym}
\usepackage{a4wide}
\usepackage{graphicx, subfigure}
\usepackage{cite}
\usepackage{tabularx}
\usepackage{array}
\usepackage{graphics}     
\usepackage{amsmath}
\usepackage{amssymb}
\usepackage{longtable} 
\usepackage{verbatim} 
\usepackage[table]{xcolor}
\usepackage{booktabs}
\usepackage{hyperref} 
\usepackage[utf8]{inputenc}
\usepackage{soul}
\usepackage[table]{xcolor}
\usepackage[normalem]{ulem}

\usepackage{pifont}

\definecolor{light-gray}{gray}{0.90}

\renewcommand{\arraystretch}{1.2} 
 
\newcommand{\cen}[1]{\multicolumn{1}{c}{#1}}
\newcommand{\ra}[1]{\renewcommand{\arraystretch}{#1}}
\newcommand{\rb}[1]{\renewcommand{\tabcolsep}{#1}}

\begin{document}

\hfill {\tt  CERN-TH-2020-093, MITP/20-030}

\def\thefootnote{\fnsymbol{footnote}}
 
\begin{center}

\vspace{3.cm}

{\Large\bf{On the new LHCb angular analysis of $B \to K^* \mu^+ \mu^-$:\\
\vspace{0.2cm}
Hadronic effects or New Physics?} }

\setlength{\textwidth}{11cm}
                    
\vspace{2.cm}
{\large\bf  
T.~Hurth$^{a,}$\footnote{Email: tobias.hurth@cern.ch},
F.~Mahmoudi$^{b,c,}$\footnote{Also at Institut Universitaire de France, 103 boulevard Saint-Michel, 75005 Paris, France}$^{,}$\footnote{Email: nazila@cern.ch},
S.~Neshatpour$^{b,}$\footnote{Email: neshatpour@ipnl.in2p3.fr }
}
 
\vspace{1.cm}
{\em $^a$PRISMA+ Cluster of Excellence and Institute for Physics (THEP)\\
Johannes Gutenberg University, D-55099 Mainz, Germany}\\[0.2cm] 
{\em $^b$Universit\'e de Lyon, Universit\'e Claude Bernard Lyon 1, CNRS/IN2P3, \\
Institut de Physique des 2 Infinis de Lyon, UMR 5822, F-69622, Villeurbanne, France}\\[0.2cm]
{\em $^c$Theoretical Physics Department, CERN, CH-1211 Geneva 23, Switzerland} \\[0.2cm]

\end{center}

\renewcommand{\thefootnote}{\arabic{footnote}}
\setcounter{footnote}{0}

\vspace{1.cm}
\thispagestyle{empty}
\centerline{\bf ABSTRACT}
\vspace{0.5cm}
The new angular analysis of the decay $B \to K^* \ell^+ \ell^-$ recently presented by the LHCb Collaboration still indicates some tensions with the Standard Model predictions.
There are several ongoing analyses to solve the problem of separating hadronic and New Physics effects in this decay, but the significance of the observed tensions in the angular observables in $B \to K^* \mu^+\mu^-$ is still dependent on a theory guesstimate of the hadronic contributions to these decays.
Using the new data from LHCb we offer two tests which make a statistical comparison to determine whether the most favoured explanation of the anomalies is New Physics or underestimated hadronic effects. We then analyse the usefulness of these tests in two future scenarios. 
Finally, we update our global fits to all available $b \to s $ data and discuss the impact of the new LHCb measurements.  
\clearpage

\section{Introduction}
In recent years the $b\to s \ell \ell $ anomalies have been among the most promising signs of physics beyond the Standard Model (SM). 
The first anomaly that showed hints for New Physics (NP) in $b\to s \ell \ell$ transitions 
was measured by LHCb in 2013 with 1~fb$^{-1}$  of data~\cite{Aaij:2013qta} in $P_5^\prime$ (for the definition, see Ref.~\cite{DescotesGenon:2012zf}) indicating NP in $C_9$. 
This tension was again confirmed by LHCb with 3~fb$^{-1}$ of data~\cite{Aaij:2015oid}, as well as by Belle and ATLAS~\cite{Abdesselam:2016llu,Aaboud:2018krd}. 
Very recently, LHCb reconfirmed the tensions in the $B \to K^* \ell^+ \ell^-$  angular observables with 4.7 fb$^{-1}$ of data~\cite{Aaij:2020nrf}. 

Further measurements by LHCb on lepton-flavour-violating observables $R_K$ and $R_{K^*}$~\cite{Aaij:2014ora,Aaij:2017vbb,Aaij:2019wad} suggested that the observed deviations can be described by a common NP effect (at the level of more than $2\sigma$) which has reinforced the NP interpretation of the anomalies. 
Hence, besides the significance of each of these tensions the coherence (or the lack of it) is a gauge of the viability of the NP interpretation.

While the anomalies in the $R_K$ and $R_{K^*}$ ratios indirectly prove the existence of NP in $P_5^\prime$, the latter does not necessarily need to violate lepton flavour. 
However, unlike the theoretically clean $R_K$ and $R_{K^*}$ ratios which have near-perfect cancellation of hadronic uncertainties,
the $B\to K^* \mu^+\mu^-$ angular observables suffer from long-distance contributions. 
Several efforts to estimate the power corrections are 
ongoing~\cite{Khodjamirian:2010vf,Khodjamirian:2012rm,Dimou:2012un,Lyon:2013gba,Bobeth:2017vxj,Chrzaszcz:2018yza,Blake:2017fyh}, but the situation is not yet completely settled.
Therefore the significance of any NP interpretation depends on the assumptions of the size of the power corrections.

Another approach is a statistical comparison of a NP fit to the data compared to a fit of a general parametrization of the unknown power corrections.   
This is possible because in several $b \to s \bar{\ell} \ell$ observables, hadronic contributions can be mimicked by NP contributions~\cite{Jager:2012uw,Jager:2014rwa} 
(especially by $C_9$ and to a lesser extent by $C_7$).
This is especially clear in the helicity amplitude description where, for example, in the $B \to K^* \ell^+ \ell^-$ decay the long-distance hadronic effect appears only in the vectorial helicity amplitude,
\begin{align}
  H_V(\lambda) &=-i\, N^\prime \Big\{ C_9^{\rm eff} \tilde{V}_{\lambda} - C_{9}'  \tilde{V}_{-\lambda}
      + \frac{m_B^2}{q^2} \Big[\frac{2\,\hat m_b}{m_B} (C_{7}^{\rm eff} \tilde{T}_{\lambda} - C_{7}'  \tilde{T}_{-\lambda})
      - 16 \pi^2 {\cal N}_\lambda \Big] \Big\} \,,
\end{align}
with ${\cal N}_\lambda(q^2) \equiv \big(\text{Leading contribution in QCDf} + h_\lambda (q^2) \big)$, 
where unknown power corrections are denoted as $h_\lambda$.
The most general ansatz for the unknown $h_\lambda$ terms respecting the analyticity of the amplitude
(up to higher-order terms in $q^2$) is given by (see Ref.~\cite{Arbey:2018ics} for more details)
\begin{align}\label{eq:hlambdapm}
  h_\pm(q^2)&= h_\pm^{(0)} + \frac{q^2}{1 \,{\rm GeV}^2}h_\pm^{(1)} + \frac{q^4}{1 \,{\rm GeV}^4}h_\pm^{(2)}\,,\\[-6pt]
\label{eq:hlambda0}
h_0(q^2)&= \sqrt{q^2}\times \left( h_0^{(0)} + \frac{q^2}{1\, {\rm GeV}^2}h_0^{(1)} + \frac{q^4}{1\, {\rm GeV}^4}h_0^{(2)}\right).
\end{align}
Instead of making assumptions about the size of the unknown power corrections, they can be directly fitted to the data~\cite{Ciuchini:2015qxb}.
While in principle, besides the $B\to K^* \ell^+ \ell^- $ decay, there are unknown power corrections
also present in the case of the $B_s\to \phi \ell^+ \ell^-$ and $B\to K \ell^+ \ell^- $ decays, there are only enough data to be able to make a meaningful hadronic fit for the former decay mode with muons (which involves 18 free parameters, considering $h_{\pm,0}^{(0,1,2)}$ to be complex).

Embedded scenarios allow us to make a statistical comparison between nested scenarios via Wilks' test.
Since the effect of the Wilson coefficients $C_{7,9}$ can in general be embedded in the general description of the unknown hadronic contributions, it is possible to also compare the hadronic fits with the NP fits by applying  Wilks' theorem. 
In light of the new data on the angular observables of the $B\to K^* \mu^+ \mu^-$ decay, we check by Wilks' test whether the updated data are constraining enough to indicate a clear preference for one scenario or the other.

Another description of hadronic contributions that can be considered as a null test for the NP explanation is via
\begin{align}
 h_\lambda (q^2)= -\frac{\tilde{V}_\lambda(q^2)}{16 \pi^2} \frac{q^2}{m_B^2}  \Delta C_9^{\lambda,\rm{PC}}\,,
\end{align}
with $\Delta C_9^{\lambda,\rm{PC}}$ being three complex (six real) $q^2$-independent parameters, this is tantamount to fitting $C_9$ with three different helicities in $H_V$.
In other words, this is a minimalistic description of power corrections where in order to rule out the NP explanation no extra $q^2$ term is needed and it will suffice if the fits to the three helicities are not compatible with each other.

Finally, we emphasize that the statistical comparison of different fits to the data can in principle only lead to indications of possible resolutions of the flavour anomalies. In general, it is true that as long as the NP fit is embedded in the more general hadronic fit, one cannot disprove the hadronic option in favour of the NP one with the set of observables considered in the present analysis. On the other hand NP can hide within hadronic contributions. Thus, NP can be firmly established by new observables such as the lepton-flavour-violating ratios mentioned above or by a real estimate of the power-suppressed terms in the SM prediction only.

This paper is organized as follows. In Section 2 we analyse the impact of the new LHCb measurements. We perform fits for New Physics in $C_9$ (and $C_7$) which in principle can mimic long-distance contributions and we also separately fit to a general parametrization of hadronic power corrections.   
We then use Wilks' test to compare both fits. Finally, we do an independent (minimal) hadronic fit which can serve as a null test for the NP option. In Section 3 we explore future prospects of these measurements and tests considering two different scenarios. 
In Section 4 we offer a new global fit to all $b \to s\ell\ell$ data using one or two operators, as well as the full set of operators. We discuss the reasons for the rather large differences compared to our previous analyses. Section 5 contains our summary.

\section{Various fits to the new data}
We consider only the exclusive $B \to K^*\, \bar\mu\mu/\gamma$ observables, namely BR$(B\to K^* \gamma)$ and the branching ratio and angular observables {$F_L, A_{\rm FB}, S_{3,4,5,7,8,9}$} of $B\to K^{*} \mu^+ \mu^-$ for the five low-$q^2$ bins ($\leqslant8$ GeV$^2$)\footnote{For the correct signs of the angular observables, see Ref.~\cite{Gratrex:2015hna}.}, and also BR($B^+\to K^{*+} \mu^+ \mu^-$) in the [1.1,6] bin  making a total of 47 observables.
The observables are calculated using SuperIso 4.1~\cite{Mahmoudi:2007vz}, giving $\chi^2_{\rm SM}=85.15$. The semileptonic and radiative SM Wilson coefficients at the $\mu_b$ scale are $C_7=-0.29, C_8=-0.16, C_9=4.20$ and $C_{10}=-4.01$.
The detailed description of our statistical methods can be found in Refs.~\cite{Hurth:2014vma,Hurth:2016fbr}.
In order to be able to make a statistical comparison we make fits to New Physics as well as to hadronic power corrections, in both cases assuming no theoretical uncertainty from long-distance contributions \cite{Chobanova:2017ghn,Arbey:2018ics} which is necessary in order to be able to apply Wilks' theorem.

We first fit the data to NP in real and complex $C_9$ and also $C_7$. 
The best-fit values as well as SM pulls are given in Table~\ref{tab:C79_set7}.

\begin{table}[h!]
\ra{1.}
\rb{1.3mm}
\begin{center}
\setlength\extrarowheight{2pt}
\scalebox{0.9}{
\begin{tabular}{|c|c|c|c|}
\hline
  \multicolumn{2}{|c}{{ $B \to K^*\, \bar\mu\mu/\gamma$ observables}} & \multicolumn{2}{c|}{($\chi^2_{\rm SM}=85.1$)  }       \\  
\hline
 & \multicolumn{1}{c|}{best-fit value} & $\chi^2_{\rm min}$ & Pull$_{\rm SM}$\\
 \hline \hline
$\delta C_9$ & $  -1.11	\pm	0.15 $ & $49.7$ &  $6.0\sigma$ \\  
\hline \hline  \rule{0pt}{3ex} 
$\delta C_7$ & $ 0.01	\pm	0.03$ & &\\[-6pt]
\footnotesize{\&} & & $49.4$ & $5.6\sigma$\\  [-4pt]
$\delta C_9$ & $ -1.22	\pm	0.25 $ & & \\
\hline
\end{tabular}  
} \quad 
\scalebox{0.9}{
\begin{tabular}{|c|c|c|c|}
\hline
  \multicolumn{2}{|c}{$B \to K^*\, \bar\mu\mu/\gamma$ observables} & \multicolumn{2}{c|}{($\chi^2_{\rm SM}=85.1$)  }       \\  
\hline
 & \multicolumn{1}{c|}{best-fit value} & $\chi^2_{\rm min}$ & Pull$_{\rm SM}$ \\
 \hline \hline
$\delta C_9$ & $(-1.04	\pm	0.17) + i(-1.24	\pm	0.61)  $ &  $47.3$ & $5.8\sigma$ \\                                                         
\hline  \hline  \rule{0pt}{3ex} 
$\delta C_7$ & $(0.01	\pm	0.03) +  i(-0.04	\pm	0.03) $ & &\\  [-6pt]
\footnotesize{\&} & & $45.6$ &  $5.4\sigma$ \\  [-4pt]
$\delta C_9$ & $(-1.15	\pm	0.28) +  i(-0.80	\pm	0.75) $ & &\\ 
\hline
\end{tabular}    
}
\caption{One- and two-operator NP fits for real (complex) $\delta C_9$ and $\delta C_{7,9}$ on the left (right), considering \mbox{$B\to K^* \bar\mu\mu/\gamma$} observables for $q^2$ bins $\leqslant 8\text{ GeV}^2$.
\label{tab:C79_set7}
}
\end{center} 
\end{table}  

In a second fit we consider the power corrections as described in Eqs.~(\ref{eq:hlambdapm}) and~(\ref{eq:hlambda0}) with 18 free parameters which leads to an improved description of the data with $4.7\sigma$ significance. 
The fitted $h_{\pm,0}^{(0,1,2)}$ parameters are given in table~\ref{tab:CompHad_set7} and, while the central values are all nonzero, within the $1\sigma$ range they are compatible with zero when taken individually, which makes it difficult to get a conclusive picture. This issue is partly due to the rather large number of degrees of freedom of the fit as well as the experimental uncertainties which are not yet small enough to give a constrained result for the fit.
In addition, we also reproduced the same fit assuming the $h_\lambda$ to be real (not shown in the table). 
With a decrease in the degrees of freedom there is an increase in the number of fitted parameters that are inconsistent with zero, however, the full $h_\lambda$ contributions for all three helicities still remain compatible with zero when taken individually.

\begin{table}[h!]
\ra{0.90}
\rb{1.3mm}
\begin{center}
\setlength\extrarowheight{2pt}
\scalebox{0.9}{
\begin{tabular}{|l||r|r|}
\hline
 \multicolumn{3}{|c|}{$B\to K^*\, \bar\mu\mu/\gamma$ observables \iffalse in the low $q^2$ bins up to 8 GeV$^2$ \fi}           \\  
 \multicolumn{3}{|c|}{ ($\chi^2_{\rm SM}=85.15,\; \chi^2_{\rm min}=25.96;\; {\rm Pull}_{\rm SM}=4.7\sigma$)}   \\ 
\hline
& \cen{Real}                         & \multicolumn{1}{c|}{Imaginary}   \\ 
 \hline
$h_{+}^{(0)}$	& $ (	-2.37	\pm	13.50	)\times 10^{-5}	$ & $ (	7.86	\pm	13.79	)\times 10^{-5}	$ \\
$h_{+}^{(1)}$	& $ (	1.09	\pm	1.81	)\times 10^{-4}	$ & $ (	1.58	\pm	1.69	)\times 10^{-4}	$ \\
$h_{+}^{(2)}$	& $ (	-1.10	\pm	2.66	)\times 10^{-5}	$ & $ (	-2.45	\pm	2.51	)\times 10^{-5}	$ \\
\hline											
$h_{-}^{(0)}$	& $ (	1.43	\pm	12.85	)\times 10^{-5}	$ & $ (	-2.34	\pm	3.09	)\times 10^{-4}	$ \\
$h_{-}^{(1)}$	& $ (	-3.99	\pm	8.11	)\times 10^{-5}	$ & $ (	1.44	\pm	2.82	)\times 10^{-4}	$ \\
$h_{-}^{(2)}$	& $ (	2.04	\pm	1.16	)\times 10^{-5}	$ & $ (	-3.25	\pm	3.98	)\times 10^{-5}	$ \\
\hline											
$h_{0}^{(0)}$	& $ (	2.38	\pm	2.43	)\times 10^{-4}	$ & $ (	5.10	\pm	3.18	)\times 10^{-4}	$ \\
$h_{0}^{(1)}$	& $ (	1.40	\pm	1.98	)\times 10^{-4}	$ & $ (	-1.66	\pm	2.41	)\times 10^{-4}	$ \\
$h_{0}^{(2)}$	& $ (	-1.57	\pm	2.43	)\times 10^{-5}	$ & $ (	3.04	\pm	29.87	)\times 10^{-6}	$ \\
\hline
\end{tabular} 
}
\caption{Hadronic power correction fit to $B\to K^*\, \bar\mu\mu/\gamma$ observables for $q^2$ bins $\leqslant 8\text{ GeV}^2$, with complex power corrections up to $q^{2\,(4)}$ terms with 18 free parameters in total. 
\label{tab:CompHad_set7}}
\end{center} 
\end{table}
The minimalistic description of hadronic corrections via $\Delta C_9^\lambda$ may in principle have a better chance to rule out the NP explanation of the data since it is described with fewer degrees of freedom. However, with the current data, as can be seen in table~\ref{tab:DeltaC9_set7}, all three helicities are still compatible with each other. As argued in Refs.~\cite{Jager:2012uw,Jager:2014rwa},  the power corrections of $H_V(\lambda=+)$ are expected to be suppressed by $\Lambda/m_B$ compared to $H_V(\lambda=-)$, and the large $\Delta C_9^+$ compared to $\Delta C_9^-$ in table~\ref{tab:DeltaC9_set7} is due the definition of $\Delta C_9^\lambda$ with a $\tilde{V}_\lambda$ factor and the fact that most of the angular observables have a minor sensitivity to $H_V(\lambda=+)$.

\begin{table}[h!]
\ra{1.}
\rb{1.3mm}
\begin{center}
\setlength\extrarowheight{2pt}
\begin{tabular}{|c|c|}
\hline
 \multicolumn{2}{|c|}{$B \to K^*\, \bar\mu\mu/\gamma$ observables}           \\  
 \multicolumn{2}{|c|}{ ($\chi^2_{\rm SM}=85.15,\; \chi^2_{\rm min}=39.40;\; {\rm Pull}_{\rm SM}=5.5\sigma$)}   \\ 
\hline
 & \multicolumn{1}{c|}{best-fit value} \\
 \hline \hline
$\Delta C_9^{+,{\rm PC}}$ & $\phantom{-}( 3.39	\pm	6.44) + i(-14.98	\pm	8.40)  $   \\                                                         
\hline 
$\Delta C_9^{-,{\rm PC}}$ & $(-1.02	\pm	0.22) + i(-0.68		\pm	0.79)  $   \\                                                         
\hline 
$\Delta C_9^{0,{\rm PC}}$ & $(-0.83	\pm	0.53) + i(-0.89		\pm	0.69)  $   \\                                                         
\hline
\end{tabular}    
\caption{Hadronic power correction fit for the three helicities ($\lambda=\pm,0$) in the form of complex $\Delta C_9^{\lambda,{\rm PC}}$, considering $B\to K^* \bar\mu\mu/\gamma$ observables for $q^2$ bins $\leqslant 8\text{ GeV}^2$.
\label{tab:DeltaC9_set7}
}
\end{center} 
\end{table}  

{While all of the aforementioned scenarios for the hadronic fits give a better description of the data compared to the SM, the current experimental data is not constraining enough to give clear individual results incompatible with zero.
Nonetheless, having nested scenarios, we can compare the different models via Wilks' test. 
In table~\ref{tab:Wilks_set7} we give the significance of the improvement of the fit when further parameters are considered.  
The results can be compared to those of table~3 in Ref. \cite{Arbey:2018ics}, but we now include several other versions of the hadronic fit. 

\begin{table}[h!]
\ra{1.2}
\rb{1.3mm}
\begin{center}
\rowcolors{4}{}{light-gray}
\setlength\extrarowheight{2pt}
\scalebox{0.8}{
\begin{tabular}{|l|c|c|c|c|c|c|c|c|c|}
\hline
\multicolumn{9}{|c|}{$B\to K^*\, \bar\mu\mu/\gamma$ observables; low-$q^2$ bins up to 8 GeV$^2$ }           \\  
\hline
\multicolumn{1}{|c|}{\multirow{2}{*}{nr. of free}} & 1 & 2 & 2 & 4 & 3  & 6 & 9  & 18 \\ [-2pt]
\multicolumn{1}{|c|}{parameters} &
$\footnotesize \left(\!\!\begin{array}{c} {\rm Real} \\ \delta C_9 \end{array}\!\!\right)$ &
$\footnotesize \left(\!\!\begin{array}{c} {\rm Real} \\ \delta C_7,\delta C_9 \end{array}\!\!\right)$ &
$\footnotesize \left(\!\!\begin{array}{c} {\rm Comp.} \\ \delta C_9 \end{array}\!\!\right)$ &
$\footnotesize \left(\!\!\begin{array}{c} {\rm Comp.} \\ \delta C_7,\delta C_9 \end{array}\!\!\right)$ &
$\footnotesize \left(\!\!\begin{array}{c} {\rm Real} \\ \Delta C_9^{\lambda,{\rm PC}} \end{array}\!\!\right)$ &
$\footnotesize \left(\!\!\begin{array}{c} {\rm Comp.} \\ \Delta C_9^{\lambda,{\rm PC}} \end{array}\!\!\right)$ &
%
$\footnotesize \left(\!\!\begin{array}{c} {\rm Real} \\ h_{+,-,0}^{(0,1,2)} \end{array}\!\!\right)$ &
$\footnotesize \left(\!\!\begin{array}{c} {\rm Comp.} \\ h_{+,-,0}^{(0,1,2)} \end{array}\!\!\right)$\\
\hline
0 (plain SM)					 	&	 $6.0\sigma$	  	&	 $5.6\sigma$ 		&	 $5.8\sigma$ 			&	 $5.4\sigma$ 		&	 $5.4\sigma$		&	 $5.5\sigma$		&	 $5.0\sigma$ 		&	 $4.7\sigma$\\
1 {\small(Real $\delta C_9$)}			 	&	 $\text{---}$	  	&	 $0.5\sigma$			&	 $1.5\sigma$ 			&	 $1.2\sigma$ 		&	 $0.6\sigma$		&	 $1.8\sigma$ 		&	 $1.1\sigma$ 		&	 $1.5\sigma$\\
2 {\small(Real $\delta C_7,\delta C_9$)}	 	&	 $\text{---}$	  	&	 $\text{---}$		&	 $\text{---}$ 			&	 $1.4\sigma$ 		&	 $\text{---}$		&	 $\text{---}$ 		&	 $1.3\sigma$ 		&	 $1.6\sigma$\\
2 {\small(Comp. $\delta C_9$)}			 	&	 $\text{---}$	  	&	 $\text{---}$		&	 $\text{---}$ 			&	 $0.8\sigma$ 		&	 $\text{---}$		&	 $1.7\sigma$ 		&	 $\text{---}$ 		&	 $1.4\sigma$\\
4 {\small(Comp. $\delta C_7,\delta C_9$)}	 	&	 $\text{---}$	  	&	 $\text{---}$ 		&	 $\text{---}$ 			&	 $\text{---}$ 		&	 $\text{---}$		&	 $\text{---}$ 		&	 $\text{---}$ 		&	 $1.5\sigma$\\
3 {\small(Real $\Delta C_9^{\lambda,{\rm PC}}$)} 	&	 $\text{---}$	  	&	 $\text{---}$ 		&	 $\text{---}$ 			&	 $\text{---}$ 		&	 $\text{---}$		&	 $2.2\sigma$ 		&	 $1.4\sigma$ 		&	 $1.7\sigma$\\
6 {\small(Comp. $\Delta C_9^{\lambda,{\rm PC}}$)}	&	 $\text{---}$	  	&	 $\text{---}$ 		&	 $\text{---}$ 			&	 $\text{---}$ 		&	 $\text{---}$		&	 $\text{---}$ 		&	 $\text{---}$ 		&	 $0.1\sigma$\\
9 {\small(Real $h_{+,-,0}^{(0,1,2)}$)}		 	&	 $\text{---}$	  	&	 $\text{---}$ 		&	 $\text{---}$ 			&	 $\text{---}$ 		&	 $\text{---}$		&	 $\text{---}$ 		&	 $\text{---}$ 		&	 $1.5\sigma$\\
\hline
\end{tabular} 
}
\caption{Improvement of fit to $B\to K^* \bar\mu\mu/\gamma$ observables for $q^2$ bins $\leqslant 8\text{ GeV}^2$, 
for the hadronic fit and the scenarios with real and complex NP  contributions to 
Wilson coefficients  $C_7$ and $C_9$ compared to the SM hypothesis and compared to each other.
\label{tab:Wilks_set7}
}
\end{center} 
\end{table}  
Considering the second row and last column in table~\ref{tab:Wilks_set7}, it can be seen that the description of the data improves with a modest significance of $1.5\sigma$ when adding 17 more parameters compared to the real NP contribution to $C_9$.  
Table~\ref{tab:Wilks_set7} gives the comparative statistical preference of various models, however, with the current experimental precision the results remain inconclusive where any preference among the various scenarios is less than $\sim2\sigma$.

It is also interesting to consider how models with complex contributions compare with corresponding models with real parameters.
In the second row it can be seen that considering a complex $C_9$ improves the description by $1.5\sigma$ compared to a real $C_9$, while from the sixth row it can be seen that a complex hadronic contribution $\Delta C_9^\lambda$ gives a $2.2\sigma$ improvement compared to a real one.  
Also the 18-parameter complex hadronic fit $h_\lambda^{(0,1,2)} $ leads to a $1.5\sigma$ improvement compared to the corresponding 9-parameter real hadronic fit.
With the considered observables both options seem to give modest improvements when complex contributions are considered and a clear distinction cannot be made.
Nonetheless, in principle CP-asymmetric observables such as $A_{7,8,9}$ can make a distinction as the imaginary parts of Wilson coefficients correspond to CP-violating ``weak'' phases, while imaginary parts in hadronic effects correspond to CP-conserving ``strong'' phases.
However, current experimental measurements of CP-asymmetric observables~\cite{Aaij:2015oid} 
do not put any constraints on the imaginary parts beyond the CP-conserving observables. 

The effect of NP contributions due to $\delta C_9$ as well as hadronic effects can be seen in Fig.~\ref{fig:S5lowq2} at the observable level for $S_5$  where the best-fit points of the fit to the real $\delta C_9$ of table~\ref{tab:C79_set7} and the hadronic fit of table~\ref{tab:CompHad_set7} are considered. 
The uncertainties of the fitted scenarios are due to theoretical uncertainties.

\vspace{0.5cm}
\begin{figure}[h!]
\centering
\includegraphics[width=0.65\textwidth]{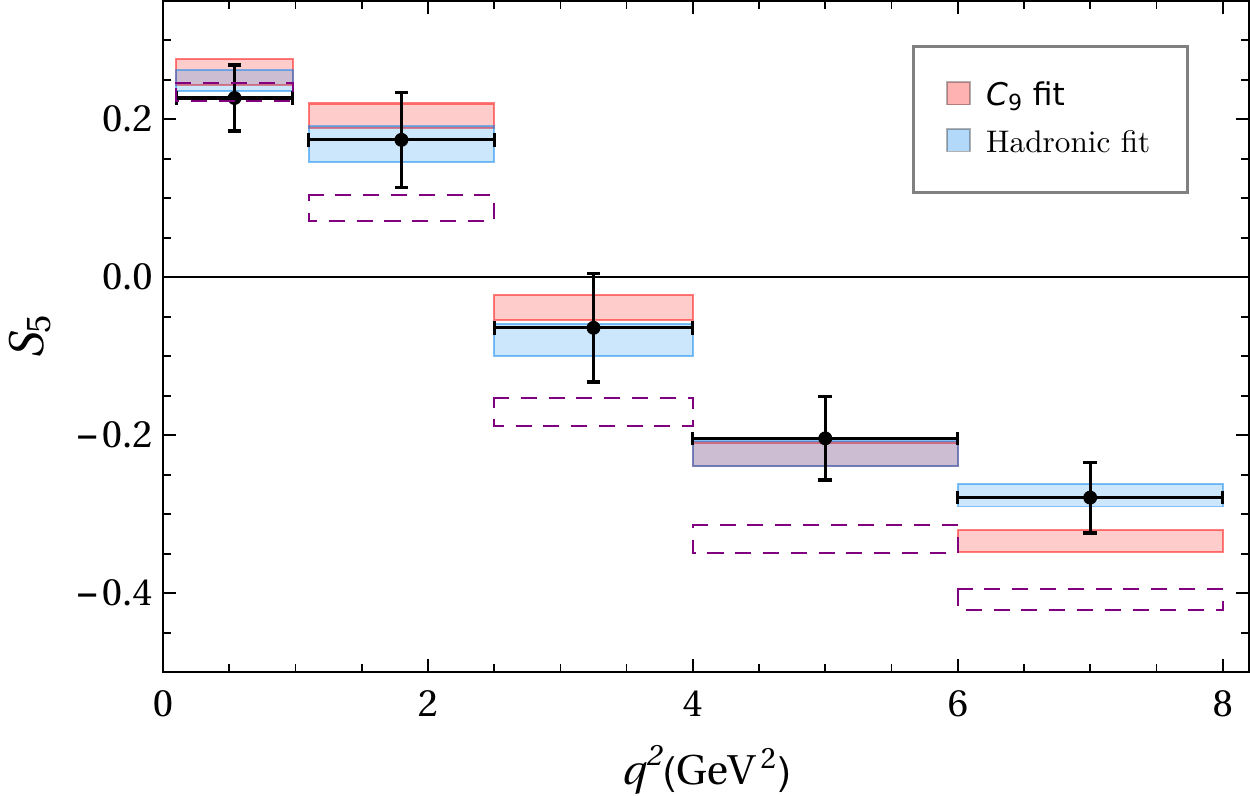}
\caption{The angular observable $S_5(B\to K^* \mu^+ \mu^-)$ with the SM predictions in dashed purple. The $C_9$ and hadronic best-fit scenarios are shown with red and blue boxes, respectively. The black crosses correspond to LHCb measurements~\cite{Aaij:2020nrf}.
\label{fig:S5lowq2}}
\end{figure}

\section{Future prospects}
We consider three benchmark points in the future: the end of Run 2 with a total integrated luminosity of 13.9\, ${\rm fb}^{-1}$~\footnote{This number corresponds to an {\it effective} luminosity of  $1\,{\rm fb}^{-1} + 2\,{\rm fb}^{-1} \times 8/7 + 5.7\,{\rm fb}^{-1} \times 13/7 \sim 13.9\, {\rm fb}^{-1}$ compared to the present {\it effective} luminosity of $1\,{\rm fb}^{-1} + 2\, {\rm fb}^{-1} \times 8/7 + 1.7\, {\rm fb}^{-1} \times 13/7 \sim 6.4\, {\rm fb}^{-1}$.}, the end of the first upgrade with 50 ${\rm fb}^{-1}$, and the end of the second upgrade at a high-luminosity LHC with 300 ${\rm fb}^{-1}$. We can assume that the statistical errors are reduced by factors $\sim 1.5$ after Run 2, $\sim 4$ after the first upgrade, and $\sim 9$ after the second upgrade~\footnote{Compared to the luminosities of both upgrades the present {\it effective} luminosity is $1\,{\rm fb}^{-1} \times 7/13  +  2\,{\rm fb}^{-1} \times 8/13 + 1.7\, {\rm fb}^{-1} \sim 3.5\, {\rm fb}^{-1}$.}. We conservatively assume that the systematic error does not get reduced by the end of Run 2. For the first upgrade we assume that the systematic errors scale with the statistical one, and also get reduced by a factor $\sim 4$. 
As discussed in Ref.~\cite{Aaij:2020nrf}, the three main sources of systematic error are the uncertainty associated with evaluating the acceptance at a fixed point in $q^2$, the biases observed when generating pseudoexperiments using the result of the best fit to data, and finally the peaking backgrounds.
For the second upgrade (HL-LHC) any consideration about the improvement of the systematic errors is highly speculative, so we conservatively assume no further reduction compared to the first upgrade.  

Having fixed the experimental uncertainties for the three benchmarks, we could keep the present central values in all three benchmark points, as is often assumed. However, it turns out that we do not get any acceptable fit with this standard assumption.
The reduced $\chi^2$ \big( $\equiv$ $\chi^2$/ (number of independent experimental observables - number of fit parameters)\big) for the NP and hadronic fits result in values much larger than 10 indicating that none of those fits describe the data correctly. 

Therefore, we use two other (equally strong) assumptions. 
In the first (hadronic) scenario, we assume that the central values of the 18 hadronic fit parameters are stable in all future benchmarks. 
In the second (New Physics) scenario, we assume that the central value of the $C_9$ parameter is always the same. 
In both scenarios the experimental central values are adjusted accordingly. Clearly, these are the two extreme assumptions when we compare NP and hadronic scenarios, and future measurements will most likely provide data that lies somewhere between these two scenarios. In the following we consider both scenarios within all three future benchmark points. We demonstrate the usefulness of the two tests to make a statistical comparison in order to find whether the most favoured explanation of the tension is New Physics or underestimated hadronic effects.

\subsection{Projections presuming $C_9$ fit central values}
Assuming that future experimental data correspond to the present best fit for the real $C_9$, one finds a perfect fit for $C_9$ at all projected benchmark points where their uncertainties are reduced for higher luminosities (see table~\ref{tab:C9projections}). The large SM pull is a very strong argument in favour of the NP scenario.
\begin{table}[h!]
\ra{1.}
\rb{1.3mm}
\begin{center}
\setlength\extrarowheight{2pt}
\scalebox{0.9}{
\begin{tabular}{|c|c|c|}
\hline
  \multicolumn{3}{|c|}{{\bf Run 2} \iffalse \;\; ($\chi^2_{\rm SM}=65.28$) \fi  }       \\  
\hline
 & \multicolumn{1}{c}{best-fit value}  & \multicolumn{1}{|c|}{Pull$_{\rm SM}$}\\
 \hline \hline
$\delta C_9$ & $	-1.11	 \pm	0.11	 $  & $8.1\sigma$ \\  
\hline
\end{tabular}  \quad
\begin{tabular}{|c|c|c|}
\hline
  \multicolumn{3}{|c|}{{\bf First LHCb upgrade} \iffalse \;\;($\chi^2_{\rm SM}=227.31$) \fi }       \\  
\hline
 & \multicolumn{1}{c}{best-fit value}  & \multicolumn{1}{|c|}{Pull$_{\rm SM}$}\\
 \hline \hline
$\delta C_9$ & $	-1.11	 \pm	0.06	 $ & $15.1\sigma$ \\  
\hline
\end{tabular}    \quad
\begin{tabular}{|c|c|c|}
\hline
  \multicolumn{3}{|c|}{{\bf HL-upgrade} \iffalse \;\; ($\chi^2_{\rm SM}=456.88$) \fi }       \\  
\hline
 & \multicolumn{1}{c}{best-fit value} &  \multicolumn{1}{|c|}{Pull$_{\rm SM}$}\\
 \hline \hline
$\delta C_9$ & $	-1.11	 \pm	0.04	 $ & $21.4\sigma$ \\  
\hline
\end{tabular} 
}
\caption{Prospect of the fit to $\delta C_9$ 
considering $B\to K^* \bar\mu\mu/\gamma$ observables for $q^2$ bins $\leqslant 8\text{ GeV}^2$, 
after Run 2 in the left, the first LHCb upgrade in the middle, and the HL-upgrade in the right table (assuming central values of the $C_9$ best fit).
\label{tab:C9projections}
}
\end{center} 
\end{table}  
Also, the two (18-parameter and 6-parameter) hadronic fits work out very well within this scenario for all three benchmarks.
This can be understood by the fact that all hadronic fits contain the $C_9$ fit. However, one finds that the uncertainties of many hadronic fit parameters become very large for higher luminosities. This fact clearly  signals that most of the 18 parameters are not needed to describe the data. Also the SM pull of the hadronic fit is always significantly smaller than the real $C_9$ fit.
The results of the 6-parameter hadronic fit are given explicitly in table~\ref{tab:DeltaC9_ProjectionsC9}. One finds that all three real parameters become more and more consistent with each other and the three imaginary parts almost vanish, while the uncertainties shrink. So this test signals that the New Physics $C_9$ fit is the favoured one. Clearly, in principle it is not possible to rule out the hadronic explanation, however, it is extremely unlikely that the power corrections for all different helicities would conspire to imitate the NP description.
Finally, Wilks' test is the one that gives a clear indication; it shows that adding additional fit parameters beyond $C_9$ does not lead to any improvement of the fit at all. 
\begin{table}[h!]
\ra{1.}
\rb{1.3mm}
\begin{center}
\setlength\extrarowheight{2pt}
\scalebox{0.82}{
\begin{tabular}{|c|c|}
\hline
 \multicolumn{2}{|c|}{\bf Run 2}           \\  [-2pt]
 \multicolumn{2}{|c|}{ (${\rm Pull}_{\rm SM}=	6.8	\sigma$)}   \\ 
\hline
 & \multicolumn{1}{c|}{best-fit value} \\
 \hline \hline
\!\!$\Delta C_9^{+,{\rm PC}}$\!\! &\!\! $(	-1.12	\pm	4.22	)\!\! +\!\! i(	-0.02	\pm	5.25	)  $\!\!   \\
\hline								
\!\!$\Delta C_9^{-,{\rm PC}}$\!\! &\!\! $(	-1.09	\pm	0.15	)\!\! +\!\! i(	-0.01	\pm	0.51	)  $\!\!   \\
\hline								
\!\!$\Delta C_9^{0,{\rm PC}}$\!\! &\!\! $(	-1.09	\pm	0.37	)\!\! +\!\! i(	-0.01	\pm	0.62	)  $\!\!   \\
\hline
\end{tabular}   
}
\scalebox{0.82}{
\begin{tabular}{|c|c|}
\hline
 \multicolumn{2}{|c|}{\bf First LHCb upgrade}           \\  [-2pt]
 \multicolumn{2}{|c|}{ (${\rm Pull}_{\rm SM}=	14.0	\sigma$)}   \\ 
\hline
 & \multicolumn{1}{c|}{best-fit value} \\
 \hline \hline
\!\!$\Delta C_9^{+,{\rm PC}}$\!\! &\!\! $(	-1.20	\pm	1.77	)\!\! +\!\! i(	0.02	\pm	1.88	)  $\!\!   \\
\hline								
\!\!$\Delta C_9^{-,{\rm PC}}$\!\! &\!\! $(	-1.09	\pm	0.07	)\!\! +\!\! i(	-0.01	\pm	0.19	)  $\!\!   \\
\hline								
\!\!$\Delta C_9^{0,{\rm PC}}$\!\! &\!\! $(	-1.10	\pm	0.16	)\!\! +\!\! i(	0.00	\pm	0.22	)  $\!\!   \\
\hline
\end{tabular}    
}
\scalebox{0.82}{
\begin{tabular}{|c|c|}
\hline
 \multicolumn{2}{|c|}{\bf HL-upgrade}           \\  [-2pt]
 \multicolumn{2}{|c|}{ (${\rm Pull}_{\rm SM}=	19.3	\sigma$)}   \\ 
\hline
 & \multicolumn{1}{c|}{best-fit value} \\
 \hline \hline
\!\!$\Delta C_9^{+,{\rm PC}}$\!\! &\!\! $(	-1.17	\pm	0.98	)\!\! +\!\! i(	0.01	\pm	0.84	)  $\!\!   \\
\hline								
\!\!$\Delta C_9^{-,{\rm PC}}$\!\! &\!\! $(	-1.09	\pm	0.05	)\!\! +\!\! i(	0.00	\pm	0.09	)  $\!\!   \\
\hline								
\!\!$\Delta C_9^{0,{\rm PC}}$\!\! &\!\! $(	-1.10	\pm	0.09	)\!\! +\!\! i(	0.00	\pm	0.10	)  $\!\!   \\
\hline
\end{tabular}    
}
\caption{Prospect of the hadronic power correction fit for
$\Delta C_9^{\lambda,{\rm PC}}$ (assuming central values of the $C_9$ best fit).
\label{tab:DeltaC9_ProjectionsC9}
}
\end{center} 
\end{table}  

\subsection{Projections presuming hadronic fit central values}

Considering the opposite extreme scenario, we presume that future data correspond to projecting the observables with the current fitted values of the 18-parameter hadronic fit.
In this case, the hadronic fit, by construction, gives a perfect fit for all benchmark points with the same central value as given in table~\ref{tab:HadronicProjections} (with small differences due to the multi-parameter fits not converging completely to the same exact values).
And their uncertainties become smaller for the higher luminosities.
Already at the end of Run 2, one notices  that 10 of the 18 fitted parameters are no longer consistent with zero, as opposed to only 2 parameters being incompatible with zero with current data (table~\ref{tab:CompHad_set7}). 
The higher the luminosities the more constrained the fit parameters, resulting in an increase in the number of parameters incompatible with zero. 
\begin{table}[h!]
\ra{0.90}
\rb{1.3mm}
\begin{center}
\setlength\extrarowheight{2pt}
\scalebox{0.9}{
\begin{tabular}{|l||r|r|}
\hline
 \multicolumn{3}{|c|}{{\bf Run 2}\;\; ($ {\rm Pull}_{\rm SM}=	7.9	\sigma$)}           \\  
\hline
& \cen{Real}                         & \multicolumn{1}{c|}{Imaginary}   \\ 
 \hline
$h_{+}^{(0)}$ & $ (	-2.39	\pm	8.75	)\times 10^{-5}	$ & $ (	7.92	\pm	8.52	)\times 10^{-5}	$ \\
$h_{+}^{(1)}$ & $ (	1.07	\pm	1.14	)\times 10^{-4}	$ & $ (	1.59	\pm	1.15	)\times 10^{-4}	$ \\
$h_{+}^{(2)}$ & $ (	-1.05	\pm	1.72	)\times 10^{-5}	$ & $ (	-2.46	\pm	1.79	)\times 10^{-5}	$ \\
\hline										
$h_{-}^{(0)}$ & $ (	2.45	\pm	10.19	)\times 10^{-5}	$ & $ (	-2.33	\pm	1.70	)\times 10^{-4}	$ \\
$h_{-}^{(1)}$ & $ (	-4.31	\pm	5.71	)\times 10^{-5}	$ & $ (	1.49	\pm	1.62	)\times 10^{-4}	$ \\
$h_{-}^{(2)}$ & $ (	2.03	\pm	0.82	)\times 10^{-5}	$ & $ (	-3.41	\pm	2.44	)\times 10^{-5}	$ \\
\hline										
$h_{0}^{(0)}$ & $ (	2.28	\pm	1.62	)\times 10^{-4}	$ & $ (	5.21	\pm	1.93	)\times 10^{-4}	$ \\
$h_{0}^{(1)}$ & $ (	1.41	\pm	1.09	)\times 10^{-4}	$ & $ (	-1.68	\pm	1.26	)\times 10^{-4}	$ \\
$h_{0}^{(2)}$ & $ (	-1.58	\pm	1.35	)\times 10^{-5}	$ & $ (	2.86	\pm	16.00	)\times 10^{-6}	$ \\
\hline
\end{tabular} \quad\quad
\begin{tabular}{|l||r|r|}
\hline
 \multicolumn{3}{|c|}{{\bf First LHCb upgrade}\;\; ($ {\rm Pull}_{\rm SM}=	22.5	\sigma$)}           \\  
\hline
& \cen{Real}                         & \multicolumn{1}{c|}{Imaginary}   \\ 
 \hline
$h_{+}^{(0)}$ & $ (	-2.44	\pm	3.17	)\times 10^{-5}	$ & $ (	8.02	\pm	3.08	)\times 10^{-5}	$ \\
$h_{+}^{(1)}$ & $ (	1.08	\pm	0.42	)\times 10^{-4}	$ & $ (	1.56	\pm	0.41	)\times 10^{-4}	$ \\
$h_{+}^{(2)}$ & $ (	-1.06	\pm	0.63	)\times 10^{-5}	$ & $ (	-2.43	\pm	0.65	)\times 10^{-5}	$ \\
\hline										
$h_{-}^{(0)}$ & $ (	2.24	\pm	6.36	)\times 10^{-5}	$ & $ (	-2.32	\pm	0.66	)\times 10^{-4}	$ \\
$h_{-}^{(1)}$ & $ (	-4.32	\pm	2.31	)\times 10^{-5}	$ & $ (	1.48	\pm	0.57	)\times 10^{-4}	$ \\
$h_{-}^{(2)}$ & $ (	2.05	\pm	0.31	)\times 10^{-5}	$ & $ (	-3.36	\pm	0.85	)\times 10^{-5}	$ \\
\hline										
$h_{0}^{(0)}$ & $ (	2.27	\pm	0.60	)\times 10^{-4}	$ & $ (	5.18	\pm	0.71	)\times 10^{-4}	$ \\
$h_{0}^{(1)}$ & $ (	1.43	\pm	0.39	)\times 10^{-4}	$ & $ (	-1.68	\pm	0.44	)\times 10^{-4}	$ \\
$h_{0}^{(2)}$ & $ (	-1.61	\pm	0.48	)\times 10^{-5}	$ & $ (	3.08	\pm	5.45	)\times 10^{-6}	$ \\
\hline
\end{tabular} 
}
\\\vspace*{0.5cm}
\scalebox{0.9}{
\begin{tabular}{|l||r|r|}
\hline
 \multicolumn{3}{|c|}{{\bf HL-upgrade}\;\; ($ {\rm Pull}_{\rm SM}=	41.8	\sigma$)}           \\  
\hline
& \cen{Real}                         & \multicolumn{1}{c|}{Imaginary}   \\ 
 \hline
$h_{+}^{(0)}$ & $ (	-2.38	\pm	1.49	)\times 10^{-5}	$ & $ (	7.95	\pm	1.41	)\times 10^{-5}	$ \\
$h_{+}^{(1)}$ & $ (	1.08	\pm	0.19	)\times 10^{-4}	$ & $ (	1.58	\pm	0.20	)\times 10^{-4}	$ \\
$h_{+}^{(2)}$ & $ (	-1.07	\pm	0.30	)\times 10^{-5}	$ & $ (	-2.45	\pm	0.33	)\times 10^{-5}	$ \\
\hline										
$h_{-}^{(0)}$ & $ (	2.10	\pm	4.52	)\times 10^{-5}	$ & $ (	-2.35	\pm	0.30	)\times 10^{-4}	$ \\
$h_{-}^{(1)}$ & $ (	-4.26	\pm	1.38	)\times 10^{-5}	$ & $ (	1.49	\pm	0.19	)\times 10^{-4}	$ \\
$h_{-}^{(2)}$ & $ (	2.04	\pm	0.18	)\times 10^{-5}	$ & $ (	-3.37	\pm	0.37	)\times 10^{-5}	$ \\
\hline										
$h_{0}^{(0)}$ & $ (	2.30	\pm	0.29	)\times 10^{-4}	$ & $ (	5.14	\pm	0.30	)\times 10^{-4}	$ \\
$h_{0}^{(1)}$ & $ (	1.42	\pm	0.19	)\times 10^{-4}	$ & $ (	-1.65	\pm	0.10	)\times 10^{-4}	$ \\
$h_{0}^{(2)}$ & $ (	-1.59	\pm	0.23	)\times 10^{-5}	$ & $ (	2.72	\pm	0.63	)\times 10^{-6}	$ \\
\hline
\end{tabular} 
}
\caption{Prospect of the hadronic power correction fit to $B\to K^* \bar\mu\mu/\gamma$ observables for $q^2$ bins $\leqslant 8\text{ GeV}^2$, after Run~2 in the upper left, the first LHCb upgrade in the upper right, and the HL-upgrade in the lower table (assuming hadronic fit central values).
\label{tab:HadronicProjections}
}
\end{center} 
\end{table}  
\begin{table}[h!]
\ra{1.}
\rb{1.3mm}
\begin{center}
\setlength\extrarowheight{2pt}
\scalebox{0.85}{
\begin{tabular}{|c|c|c|c|}
\hline
  \multicolumn{4}{|c|}{\bf {Run 2 }}       \\  
   \multicolumn{4}{|c|}{ \phantom{(\iffalse $\chi^2_{\rm SM}=	112.99	,\;$ \fi$ \chi^2_{\rm min}=	27.02	;\; {\rm Pull}_{\rm SM}=	8.2	\sigma$)}}   \\ 
\hline
 & \multicolumn{1}{c}{best-fit value} & \multicolumn{1}{|c|}{$\chi^2_{\rm min}$} & \multicolumn{1}{c|}{Pull$_{\rm SM}$}\\ 
 \hline \hline
\rule{0pt}{3ex}Real $\delta C_9$ & $	-1.11	 \pm	0.11	 $ & $	49.27	$ &  $7.9\sigma$ \\ [3pt] 
\hline \hline
\rule{0pt}{3ex}Comp. $\delta C_9$ & $(	-1.09	\pm	0.12	) + i(	-0.62	 \pm	0.71	)  $ &  $	48.70	$  &  $7.6\sigma$\\ [4pt] 
\hline
\end{tabular}    
}\quad
\scalebox{0.85}{
\begin{tabular}{|c|c|}
\hline
 \multicolumn{2}{|c|}{\bf {Run 2 }}           \\  
 \multicolumn{2}{|c|}{ (\iffalse $\chi^2_{\rm SM}=	112.99	,\;$ \fi$ \chi^2_{\rm min}=	27.34	;\; {\rm Pull}_{\rm SM}=	8.1	\sigma$)}   \\ 
\hline
 & \multicolumn{1}{c|}{best-fit value} \\
 \hline \hline
$\Delta C_9^{+,{\rm PC}}$ & $(	4.27	\pm	4.63	) + i(	-15.39	\pm	5.39	)  $   \\
\hline								
$\Delta C_9^{-,{\rm PC}}$ & $(	-0.98	\pm	0.17	) + i(	-0.57	\pm	0.48	)  $   \\
\hline								
$\Delta C_9^{0,{\rm PC}}$ & $(	-0.88	\pm	0.39	) + i(	-0.74	\pm	0.50	)  $   \\
\hline
\end{tabular}  
} 
\caption{Prospects for real or complex NP fits to $C_9$ and the hadronic power correction fit for the three helicities in the form of complex $\Delta C_9^{\lambda,{\rm PC}}$ after Run 2 (assuming hadronic fit central values).
\label{tab:C9_DeltaC9_hadr_projections}
}
\end{center} 
\end{table}  

The projected data after Run 2 give an acceptable fit for NP in $C_9$ with $\sim 8\sigma$ significance improvement compared to the SM (table~\ref{tab:C9_DeltaC9_hadr_projections}). Similar improvements are also found for the hadronic model, as can be seen for the 18-parameter fit on the upper left side of table~\ref{tab:HadronicProjections}, and also for the simpler hadronic description of $\Delta C_9^\lambda$.
Interestingly, the latter model gives slightly incompatible values for the imaginary part of the different helicities (right-hand side of table~\ref{tab:C9_DeltaC9_hadr_projections}) indicating that the NP description of the data is not completely viable.
From the Wilks' test of table~\ref{tab:Wilks_hadr_projectionsCompact1} after Run 2, the $\Delta C_9^\lambda$ hadronic model and the 18-parameter one both give a better description of the data compared to real or complex NP in $C_9$ with a significance of $\sim3.5\sigma$ and $\sim4\sigma$, respectively. However, the situation still remains inconclusive after Run~2. 

For higher luminosities, both with the first LHCb and the HL upgrades, one finds that only the 18-parameter hadronic description gives an acceptable fit.
The other models -- while formally describing the data better compared to the SM with around $\sim 15\sigma$ significance (see table~\ref{tab:Wilks_hadr_projectionsCompact1}) --
do not lead to acceptable fits as they all have very large reduced $\chi^2$ statistics with small $p$-values $\simeq0$. 
With the upgraded LHCb data, unlike the  Run 2 data, the improvement of the hadronic fit compared to the SM and the NP fit is now very significant and it is possible to make a conclusive judgment regarding the preference of the hadronic description compared to NP.

\begin{table}[t!]
\ra{1.2}
\rb{1.3mm}
\begin{center}
\setlength\extrarowheight{2pt}
\scalebox{0.75}{
\rowcolors{5}{light-gray}{}
\begin{tabular}{|l|c|c|c|c|c|}
\hline
\multicolumn{5}{|c|}{{\bf Run 2} }           \\  
\hline
\multicolumn{1}{|c|}{\multirow{2}{*}{nr. of free}} & 1 & 2 & 6 & 18 \\ [-2pt]
\multicolumn{1}{|c|}{parameters} &
$\scriptsize \left(\!\!\begin{array}{c} {\rm Real} \\ \delta C_9 \end{array}\!\!\right)$ &
$\scriptsize \left(\!\!\begin{array}{c} {\rm Comp.} \\ \delta C_9 \end{array}\!\!\right)$ &
$\scriptsize \left(\!\!\begin{array}{c} {\rm Comp.} \\ \Delta C_9^{\lambda,{\rm PC}} \end{array}\!\!\right)$ &
$\scriptsize \left(\!\!\begin{array}{c} {\rm Comp.} \\ h_{+,-,0}^{(0,1,2)} \end{array}\!\!\right)$\\
\hline
0 (plain SM) 	&	 $7.9\sigma$  	&	 $7.6\sigma$ 	&	 $8.1\sigma$	&	 $7.9\sigma$\\
1 {\footnotesize(Real \!$\delta C_9$)} 	&	 $\text{---}$  	&	 $0.8\sigma$ 	&	 $3.5\sigma$ 	&	 $4.0\sigma$\\
2 {\footnotesize(Comp. \!$\delta C_9$)} 	&	 $\text{---}$  	&	 $\text{---}$ 	&	 $3.6\sigma$ 	&	 $4.1\sigma$\\
6 {\footnotesize(Comp. \!$\Delta C_9^{\lambda,{\rm PC}}$)}	&	 $\text{---}$  	&	 $\text{---}$ 	&	 $\text{---}$ 	&	 $2.7\sigma$\\
\hline
\end{tabular} 
}
\quad
\scalebox{0.75}{
\rowcolors{5}{light-gray}{}
\begin{tabular}{|l|c|c|c|c|c|}
\hline
\multicolumn{5}{|c|}{{\bf First LHCb upgrade} }           \\  
\hline
\multicolumn{1}{|c|}{\multirow{2}{*}{nr. of free}} & 1 & 2 & 6 & 18 \\ [-2pt]
\multicolumn{1}{|c|}{parameters} &
$\scriptsize \left(\!\!\begin{array}{c} {\rm Real} \\ \delta C_9 \end{array}\!\!\right)$ &
$\scriptsize \left(\!\!\begin{array}{c} {\rm Comp.} \\ \delta C_9 \end{array}\!\!\right)$ &
$\scriptsize \left(\!\!\begin{array}{c} {\rm Comp.} \\ \Delta C_9^{\lambda,{\rm PC}} \end{array}\!\!\right)$ &
$\scriptsize \left(\!\!\begin{array}{c} {\rm Comp.} \\ h_{+,-,0}^{(0,1,2)} \end{array}\!\!\right)$\\
\hline
0 (plain SM) 	&	 $14.6\sigma$  	&	 $14.4\sigma$ 	&	 $18.7\sigma$	&	 $22.5\sigma$\\
1 {\footnotesize(Real \!$\delta C_9$)} 	&	 $\text{---}$  	&	 $0.7\sigma$ 	&	 $12.0\sigma$ 	&	 $17.5\sigma$\\
2 {\footnotesize(Comp.\! $\delta C_9$)} 	&	 $\text{---}$  	&	 $\text{---}$ 	&	 $12.1\sigma$ 	&	 $17.6\sigma$\\
6 {\footnotesize(Comp. \!$\Delta C_9^{\lambda,{\rm PC}}$)}	&	 $\text{---}$  	&	 $\text{---}$ 	&	 $\text{---}$ 	&	 $12.9\sigma$\\
\hline
\end{tabular} 
}
\\ [10pt]
\scalebox{0.75}{
\rowcolors{5}{light-gray}{}
\begin{tabular}{|l|c|c|c|c|c|}
\hline
\multicolumn{5}{|c|}{{\bf HL-upgrade}}           \\  
\hline
\multicolumn{1}{|c|}{\multirow{2}{*}{nr. of free}} & 1 & 2 & 6 & 18 \\ [-2pt]
\multicolumn{1}{|c|}{parameters} &
$\scriptsize \left(\!\!\begin{array}{c} {\rm Real} \\ \delta C_9 \end{array}\!\!\right)$ &
$\scriptsize \left(\!\!\begin{array}{c} {\rm Comp.} \\ \delta C_9 \end{array}\!\!\right)$ &
$\scriptsize \!\!\left(\!\!\begin{array}{c} {\rm Comp.}\!\! \\ \Delta C_9^{\lambda,{\rm PC}} \end{array}\!\!\right)$ &
$\scriptsize \left(\!\!\begin{array}{c} {\rm Comp.} \\ h_{+,-,0}^{(0,1,2)} \end{array}\!\!\right)$\\
\hline
0 (plain SM) 	\!&	 $18.9\sigma$  \!	&	 $18.8\sigma$ 	&	 $32.7\sigma$	&	 $41.8\sigma$\\
1 {\footnotesize(Real \!$\delta C_9$)} 	\!&	 $\text{---}$  	&	 $0.7\sigma$ 	&	 $26.8\sigma$ 	&	 $37.4\sigma$\\
2 {\footnotesize(Comp. \!$\delta C_9$)} 	\!&	 $\text{---}$  	&	 $\text{---}$ 	&	 $26.9\sigma$ 	&	 $37.4\sigma$\\
6 {\footnotesize(Comp. \!$\Delta C_9^{\lambda,{\rm PC}}$)}	\!&	 $\text{---}$  	&	 $\text{---}$ 	&	 $\text{---}$ 	&	 $26.2\sigma$\\
\hline
\end{tabular} 
}
\caption{Prospect of improving the fit to $B\to K^* \bar\mu\mu/\gamma$ observables for $q^2$ bins $\leqslant 8\text{ GeV}^2$, after Run 2 in the upper left table, the first LHCb upgrade in the upper right table, and the HL-upgrade in the lower table (assuming hadronic fit central values), for the hadronic fit and the scenarios with real and complex NP contributions to the Wilson coefficient $C_9$ compared to the SM hypothesis and compared to each other.
\label{tab:Wilks_hadr_projectionsCompact1}
}
\end{center} 
\end{table}  
\vspace{0.5cm}
 
The hadronic fit and the corresponding 68\% confidence level region for the current data (table~\ref{tab:CompHad_set7}) as well as for the higher luminosities (table~\ref{tab:HadronicProjections}) can be seen in Fig.~\ref{fig:h_lambda} for Re($h_\lambda$).
Compared to the tables where only the individual uncertainties are reported, in the plots their correlations are also taken into account.
One finds again that only after the first LHCb upgrade the fitted parameters become individually inconsistent with zero at the $1\sigma$ level.
From Fig.~\ref{fig:h_lambda}, the size of the hadronic fit can be directly compared to the leading-order QCDf calculated contributions by considering the black and red solid lines, respectively. 
\begin{figure}[!th]
\begin{center}
\includegraphics[width=0.43\textwidth]{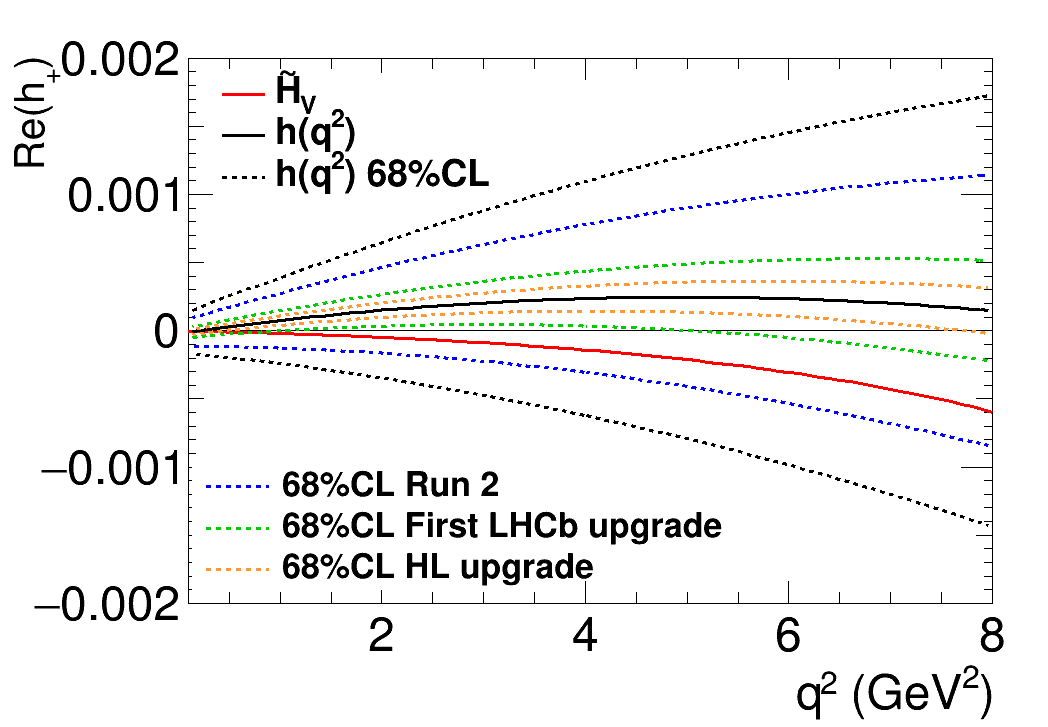}
\quad\quad\includegraphics[width=0.43\textwidth]{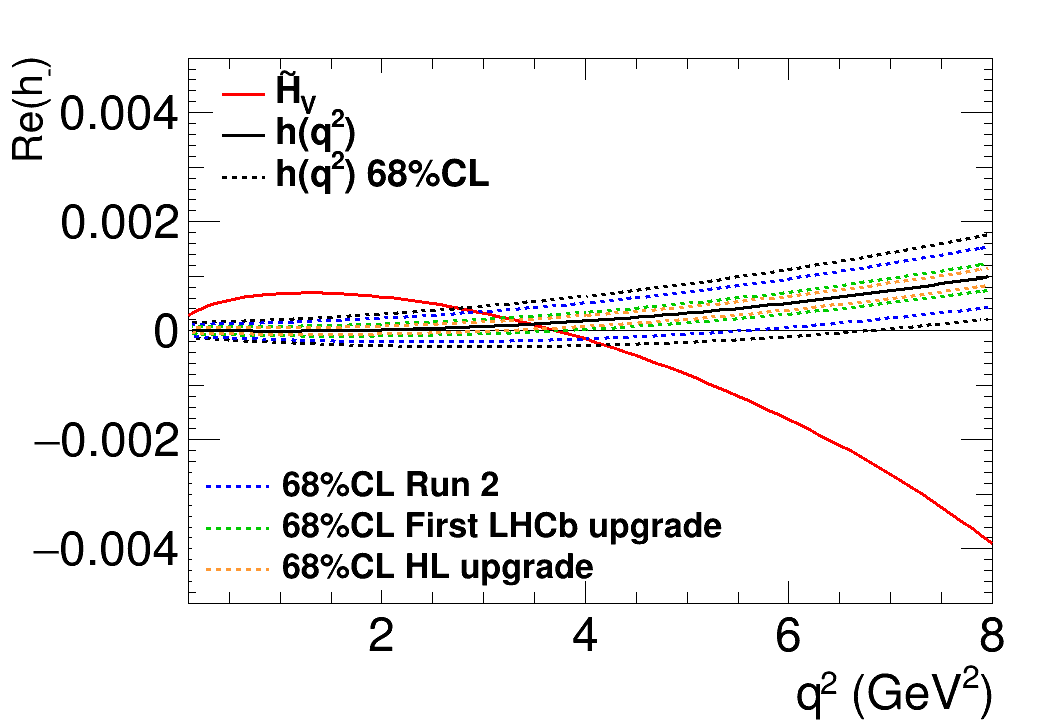}
\includegraphics[width=0.43\textwidth]{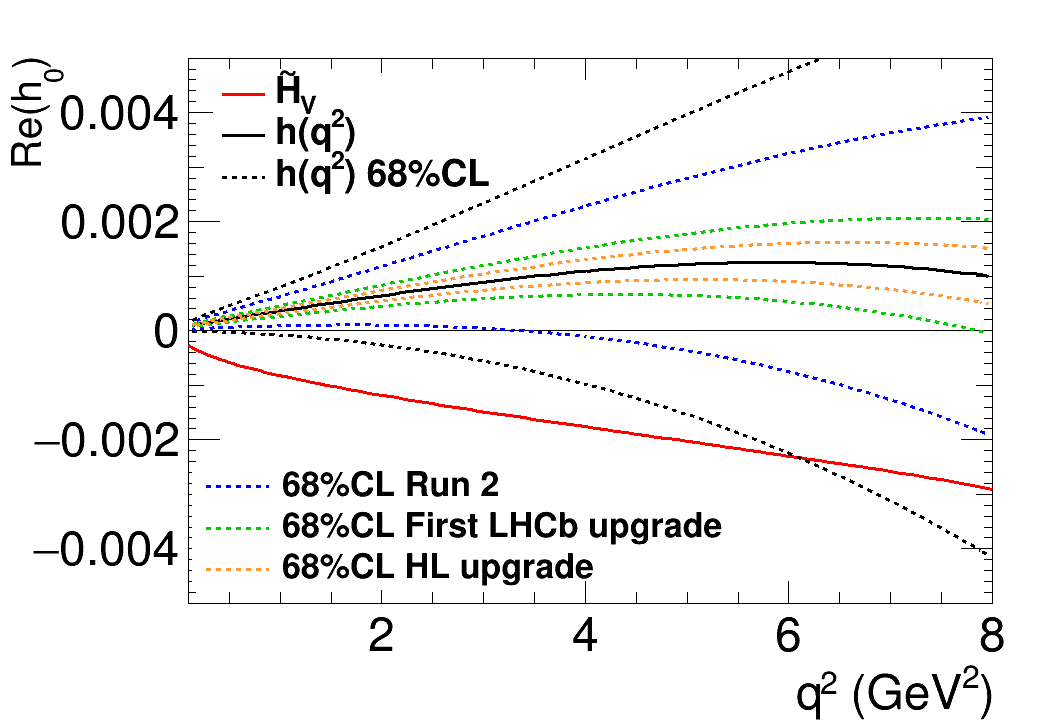}
\caption{Hadronic contributions to the helicity amplitudes.
The red line corresponds to the leading-order QCDf contribution. The extra tilde sign indicates that the helicity amplitude $\tilde{H}_V$ has been normalized by $(i16\pi^2 N^\prime m_B^2/q^2)$ to be directly comparable to the power corrections $h_\lambda$.
The solid black line corresponds to the best-fit solution of the hadronic fit.
The dashed black lines indicate the 68\% C.L. region of the hadronic fit parameter uncertainties with current LHCb data and the dashed blue, green, and yellow lines show the 68\% C.L. region for Run 2, the first LHCb upgrade, and the HL-upgrade, respectively.
\label{fig:h_lambda}}
\end{center}
\end{figure}

\section{NP fit to all $b\to s$ observables}
We present our new global fits considering the updated measurement of $B\to K^* \mu^+ \mu^-$ angular observables. Compared to our previous work~\cite{Arbey:2019duh}, we have also updated the upper bound on $B_s \to e^+ e^-$ from Ref.~\cite{Aaij:2020nol} which is now more than one order of magnitude lower compared to the previous experimental bound. 
Moreover, we consider the angular observables and branching ratio of the semileptonic baryonic decay $\Lambda_b \to \Lambda \mu^+ \mu^-$~\cite{Aaij:2015xza,Aaij:2018gwm,Detmold:2016pkz,Meinel:2016grj,Blake:2017une} resulting in 117 observables overall. As usual, we have now considered 10\% power correction uncertainty for the $B\to K^{(*)} \mu^+ \mu^-$ and $B_s \to \phi \mu^+ \mu^-$ decays within our SM predictions. Their concrete parametrization, implemented in SuperIso~\cite{Mahmoudi:2007vz}, is consistent with the most general ansatz for the hadronic power corrections compatible with analyticity (see Eq.~(\ref{eq:hlambdapm})).

We guide the reader to Refs.~\cite{Hurth:2017hxg,Arbey:2018ics,Arbey:2019duh,Alguero:2019ptt,Alok:2019ufo,Ciuchini:2019usw,Datta:2019zca,Aebischer:2019mlg,Kowalska:2019ley} for previous model-independent analyses of $b\to s \ell \ell$ data as well as to Refs.~\cite{Alguero:2019ptt,Biswas:2020uaq,Bhom:2020lmk} which considered the recent LHCb update. 

The one- and two-dimensional global fits of the Wilson coefficients to all observables are given in tables~\ref{tab:all_1D} and \ref{tab:all_2D}, respectively.
From table~\ref{tab:all_1D}, it can be seen that for the most prominent NP scenarios we have a more than $1\sigma$ increase compared to our previous fit which did not include the new LHCb update of the angular observables.

\begin{table}[th!]
\begin{center}
\setlength\extrarowheight{0pt}
\hspace*{-1.cm} 
\scalebox{0.80}{
\begin{tabular}{|l|r|r|c|}
\hline 
 \multicolumn{4}{|c|}{\small All observables  ($\chi^2_{\rm SM}=157.3$)} \\ \hline
                          & b.f. value & $\chi^2_{\rm min}$ & ${\rm Pull}_{\rm SM}$  \\ 
\hline \hline
$\delta C_{9} $    		& $ 	-0.94	\pm	0.14	 $ & $ 	126.8	 $ & $	5.5 	\sigma	 $  \\
$\delta C_{9}^{\mu} $    	& $ 	-0.93	\pm	0.13	 $ & $ 	115.2	 $ & $	6.5	\sigma	 $  \\
$\delta C_{9}^{e} $    		& $ 	0.84	\pm	0.26	 $ & $ 	145.5	 $ & $	3.4	\sigma	 $  \\
\hline
$\delta C_{10} $    		& $ 	0.20	\pm	0.22	 $ & $ 	156.4	 $ & $	0.9	\sigma	 $  \\
$\delta C_{10}^{\mu} $    	& $ 	0.51	\pm	0.17	 $ & $ 	146.4	 $ & $	3.3	\sigma	 $  \\
$\delta C_{10}^{e} $    	& $ 	-0.78	\pm	0.23	 $ & $ 	144.3	 $ & $	3.6	\sigma	 $  \\
\hline
$\delta C_{\rm LL}^\mu$ 	& $ 	-0.53	\pm	0.10	 $ & $ 	125.4	 $ & $	5.6	\sigma	 $  \\
$\delta C_{\rm LL}^e$ 		& $ 	0.43	\pm	0.13	 $ & $ 	144.8	 $ & $	3.5	\sigma	 $  \\
\hline
\end{tabular}
} 
\caption{One-operator NP fit to all $b\to s$ transitions, assuming 10\% error for the power corrections.
\label{tab:all_1D}} 
\end{center} 
\end{table}

\begin{table}[th!]
\begin{center}
\setlength\extrarowheight{3pt}
\scalebox{0.8}{
\begin{tabular}{|c|c|c|c|}
\hline 
 \multicolumn{4}{|c|}{All observables  ($\chi^2_{\rm SM}=157.3$)} \\ \hline
                          & b.f. value & $\chi^2_{\rm min}$ & ${\rm Pull}_{\rm SM}$ \\ 
\hline \hline
$\{ \delta C_{9}^{\mu} , \delta C_{9}^{\prime \mu} \}$		& $\{ 	-0.97	\pm	0.12	 \;,\;	0.34	\pm	0.20	\}$ & $	112.25	$ & $	6.4	\sigma	 $ \\
$\{ \delta C_{9}^{\mu} , \delta C_{9}^{e} \}$			& $\{ 	-0.93	\pm	0.14	 \;,\;	-0.01	\pm	0.31	\}$ & $	115.19	$ & $	6.2	\sigma	 $ \\
$\{ \delta C_{9}^{\mu} , \delta C_{10}^{\mu} \}$		& $\{ 	-0.91	\pm	0.13	 \;,\;	0.13	\pm	0.15	\}$ & $	114.41	$ & $	6.2	\sigma	 $ \\
$\{ \delta C_{LL}^{\mu} , \delta C_{LL}^{e} \}$			& $\{ 	-0.67	\pm	0.15	 \;,\;	-0.26	\pm	0.20	\}$ & $	123.74	$ & $	5.4	\sigma	 $ \\
$\{ \delta C_{LR}^{\mu} , \delta C_{LR}^{e} \}$			& $\{ 	-0.46	\pm	0.12	 \;,\;	-1.75	\pm	0.28	\}$ & $	131.59	$ & $	4.7	\sigma	 $ \\
$\{ \delta C_{9} , \delta C_{LL}^{\mu} \}$			& $\{ 	-0.64	\pm	0.17	 \;,\;	-0.34	\pm	0.10	\}$ & $	113.91	$ & $	6.3	\sigma	 $ \\
$\{ \delta C_{9} , \delta C_{LL}^{e} \}$			& $\{ 	-0.94	\pm	0.14	 \;,\;	0.42	\pm	0.13	\}$ & $	113.96	$ & $	6.3	\sigma	 $ \\
\hline
\end{tabular}
}
\caption{Two-operator NP fit to all $b\to s$ transitions, assuming 10\% error for the power corrections.}
\label{tab:all_2D} 
\end{center} 
\end{table}

We also consider a multi-dimensional NP fit consisting of 20 Wilson coefficients in total. The results are given in table~\ref{tab:ALL_20D_C78910C12primes}.
In our previous fits some of them remained undetermined due to their large uncertainties. This is no longer the case with the new data. 
One of the reasons is that we now have more significant $B_s \to e e $ data which were not available in the previous analysis.
In this global fit we also find a significant increase ($0.8\sigma$) in the SM pull compared to the previous global fit using 20 Wilson coefficients (see table 8 in Ref.~\cite{Arbey:2018ics}).
\begin{table}[th!]
\begin{center}
\setlength\extrarowheight{3pt}
\scalebox{0.78}{
\begin{tabular}{|c|c|c|c|}
\hline 
  \multicolumn{4}{|c|}{All observables with $\chi^2_{\rm SM}=157.28$} \\ 
  \multicolumn{4}{|c|}{($\chi^2_{\rm min}=100.34;\; {\rm Pull}_{\rm SM}=4.3\sigma$)} \\ 
\hline \hline
\multicolumn{2}{|c}{$\delta C_7$} &  \multicolumn{2}{|c|}{$\delta C_8$}\\ 
\multicolumn{2}{|c}{	$	0.05	\pm	0.03	$} & \multicolumn{2}{|c|}{	$	-0.71	\pm	0.43	$}\\
\hline 
\multicolumn{2}{|c}{$\delta C_7^\prime$} &  \multicolumn{2}{|c|}{$\delta C_8^\prime$}\\ 
\multicolumn{2}{|c}{	$	-0.01	\pm	0.02	$} & \multicolumn{2}{|c|}{	$	-0.09	\pm	0.86	$}\\
\hline \hline
$\delta C_{9}^{\mu}$ & $\delta C_{9}^{e}$ & $\delta C_{10}^{\mu}$ & $\delta C_{10}^{e}$ \\
$	-1.11	\pm	0.19	$  & $	-6.69	\pm	1.37	$  & $	0.08	\pm	0.25	$  & $	3.97	\pm	4.99	$ \\
\hline 
$\delta C_{9}^{\prime \mu}$ & $\delta C_{9}^{\prime e}$ & $\delta C_{10}^{\prime \mu}$ & $\delta C_{10}^{\prime e}$ \\
$	0.18	\pm	0.35	$  & $	1.84	\pm	1.75	$  & $	-0.13	\pm	0.21	$  & $	0.05	\pm	5.01	$ \\
\hline \hline
$C_{Q_{1}}^{\mu}$ & $C_{Q_{1}}^{e}$ & $C_{Q_{2}}^{\mu}$ & $C_{Q_{2}}^{e}$ \\ 
$	-0.07	\pm	0.12	$  & $	-1.52	\pm	0.98	$  & $	-0.10	\pm	0.14	$  & $	-4.36	\pm	1.46	$ \\
\hline 
$C_{Q_{1}}^{\prime \mu}$ & $C_{Q_{1}}^{\prime e}$ & $C_{Q_{2}}^{\prime \mu}$ & $C_{Q_{2}}^{\prime e}$ \\ 
$	0.05	\pm	0.12	$  & $	-1.40	\pm	1.56	$  & $	-0.17	\pm	0.15	$  & $	-4.33	\pm	2.33	$ \\
\hline
\end{tabular}
}
\caption{Best-fit values for the 20-operator fit to all observables, assuming $10\%$ error for the power corrections.
Previously, we found a SM pull of $3.5\sigma$ (see table 8 in Ref.~\cite{Arbey:2018ics}).
\label{tab:ALL_20D_C78910C12primes}} 
\end{center} 
\end{table}
%

The question arises of whether the large increase of the NP significance in all global fits can be traced back to the new LHCb data on the angular observables or if the increase just indicates that the various tensions within the $b \to s$ data are now more coherent. To resolve this question we make a fit to $\delta C_9$, considering the new data only, namely, the $B \to K^* \mu^+\mu^-$ angular observables in the low- and high-$q^2$ bins. Without adding any uncertainty for possible hadronic power corrections within the SM predictions, we find a $5.5\sigma$ NP significance which has to be compared to the $3.9\sigma$ of our previous fit to the same set of observables. 
These results clearly show that the new LHCb data on the angular observables\cite{Aaij:2020nrf} is the source of this large increase of the NP significance in all of our new global fits. 
This can be understood by the significantly larger $\chi_{\rm SM}^2$ of the new LHCb data ($\sim\!24$ units larger compared to previous data) which is due to smaller experimental uncertainties as well as the emergence of further local tensions (e.g. in $S_3([1.1,2.5])$ and $A_{FB}([6,8])$). 
The SM predictions of SuperIso lead to slightly smaller NP significance because of the $10\%$  guesstimate of power corrections included in our final SM predictions. 

The $5.5\sigma$ NP significance of the present SM predictions, in which no uncertainties due to the power corrections were added, can now also be directly compared with the $3.3\sigma$ significance that LHCb found in their recent analysis of the same set of observables~\cite{Aaij:2020nrf} using the SM predictions based on the Flavio package~\cite{Straub:2018kue}. 
Comparing the SM predictions in SuperIso and Flavio, we find that the same set of form factors is used which were calculated using the QCD sum rule approach~\cite{Straub:2015ica}. By cross-checking some central values of observables we do not find any significant differences. The parametrizations of the unknown power corrections are similar and also consistent with the analyticity constraint (see Eq.~(\ref{eq:hlambdapm})) in both cases. However, the concrete numbers chosen within the parametrizations are obviously different: larger values are assumed (guesstimated) in Flavio than in SuperIso, resulting in more conservative predictions of the unknown power corrections and a much lower NP significance. Clearly these guesstimates are guided by many concrete theoretical analyses on these unknown power corrections~\cite{Khodjamirian:2010vf,Khodjamirian:2012rm,Bobeth:2017vxj,Chrzaszcz:2018yza,Blake:2017fyh} (see Ref.~\cite{Arbey:2018ics} for a brief discussion), but a real estimate of the hadronic power corrections has not yet been established. 

Finally, we emphasize that all of our tests presented in the previous sections do not rely on any guesstimate of the unknown hadronic power corrections, but instead represent a statistical comparison of NP and hadronic fits to find indications of whether the most favoured explanation of the tensions in the $b \to s$ data is NP or underestimated hadronic power corrections. 

\section{Summary}
We analysed the recent data on the angular observables of the decay $B \to K^* \ell^+ \ell^-$. There still are some tensions with the SM predictions. 
In contrast to the theoretically clean $R_K$ and $R_{K^*}$ ratios, the $B\to K^* \mu^+\mu^-$ angular observables suffer from unknown long-distance contributions. 
Several efforts to estimate these power corrections are ongoing, but a real estimate has not been established yet. Thus, the significance of the observed tension still depends on the theoretical assumptions on the size of the power corrections.
We offered two statistical tests that allowed us to find indications of whether the tensions in the angular observables are signs of New Physics or just due to underestimated hadronic corrections. These tests do not rely on any guesstimate of the unknown power corrections but represent a statistical comparison of the NP fit and hadronic fits using the most general parametrization of the unknown power corrections compatible with analyticity. We have shown the usefulness of these tests in two different scenarios using three future benchmarks: at end of the Run 2 with a total integrated luminosity of 13.9\, ${\rm fb}^{-1}$, at the end of the first LHCb upgrade with 50 ${\rm fb}^{-1}$, and at the end of the second upgrade at a high-luminosity LHC with 300 ${\rm fb}^{-1}$. 
In addition, we updated our global fits to all $b \to s \ell\ell$ data using one or two operators and also the full set of operators.
We found an increase of the NP significance of these fits by around  $1 \sigma$. This large increase can be traced back to the new LHCb measurements of the angular observables.


\section*{Acknowledgements}
The work of T.H. was supported by  the  Cluster of  Excellence ``Precision  Physics, Fundamental Interactions, and Structure of Matter" (PRISMA$^+$ EXC 2118/1) funded by the German Research Foundation (DFG) within the German Excellence Strategy (Project ID 39083149), as well as BMBF Verbundprojekt 05H2018 - Belle II.  
S.N. has received funding from the European Union’s Horizon 2020 research and innovation programme under the Marie Sklodowska-Curie grant agreement No. 674896 and No. 690575.
TH thanks the CERN theory group for its hospitality during his regular visits to CERN where part of the work was done.

\providecommand{\href}[2]{#2}\begingroup\raggedright
\endgroup

\end{document}